\documentclass[preprint2]{aastex6}

\usepackage{xspace}
\usepackage{hyperref}

\shorttitle{DDO51 Halo Giants}
\shortauthors{Slater et al.}

\slugcomment{}

\begin{document}

\newcommand{\gddo}{$\Delta(g - \rm{DDO51})$}

\title{The Stellar Density Profile of the Distant Galactic Halo}

\author{Colin T. Slater\altaffilmark{1},
David L. Nidever\altaffilmark{2,3,4},
Jeffrey A. Munn\altaffilmark{5},
Eric F. Bell\altaffilmark{4},
Steven R. Majewski\altaffilmark{6}}
\affil{}

\altaffiltext{1}{Department of Astronomy, University of Washington, Box 351580,
Seattle, WA 98195, USA; \href{mailto:ctslater@uw.edu}{ctslater@uw.edu}}

\altaffiltext{2}{Large Synoptic Survey Telescope, 950 North Cherry Ave, Tucson,
AZ 85719; \href{mailto:dnidever@lsst.org}{dnidever@lsst.org}}

\altaffiltext{3}{Steward Observatory, 933 North Cherry Ave, Tucson, AZ 85719}

\altaffiltext{4}{Department of Astronomy, University of Michigan,
    1085 S. University Ave., Ann Arbor, MI 48109;}

\altaffiltext{5}{US Naval Observatory, Flagstaff Station, P.O. Box 1149,
Flagstaff, AZ 86005}

\altaffiltext{6}{Department of Astronomy, University of Virginia,
Charlottesville, VA 22904}

\begin{abstract}
We use extensive gravity-sensitive DDO 51 photometry over 5100 square degrees,
combined with SDSS broadband photometry, to select a catalog of $\sim 4,000$ giant
stars covering a large fraction of the high Galactic latitude sky and reaching
out to $\sim 80$ kpc in the Galactic halo. This sample of bright and unbiased
tracers enables us to measure the radial profile and 3D structure of the stellar
halo to large distance which had previously only been measured with sparse
tracers or small samples. Using population synthesis models to reproduce the
observed giant star luminosity function, we find that the halo maintains a
$r^{-3.5}$ profile from $30$ to $80$ kpc with no signs of a truncation or sharp
break over this range. The radial profile measurement is largely
insensitive to individual halo substructure components, but we find that attempting
to measure the shape of the halo is overwhelmed by the Sagittarius
stream such that no ellipsoidal shape is a satisfactory description in this
region. These measurements allow us to begin placing the Milky Way in context
with the growing sample of external galaxies where similar halo profile
measurements are available, with the goal of further linking the properties of
stellar halos to the accretion histories that formed them.
\end{abstract}


\section{Introduction}

The growth of stellar halos by accretion of satellite galaxies is a well-
established feature of galaxy evolution. A wealth of individual structures are
readily identifiable in large-scale surveys of the halo
\citep[e.g.,][]{majewski03,belokurov06,slater14}, and statistical studies of
fluctuations in number counts agree well with simulations of halos formed by
numerous accretion events \citep{bell08,xue11}. While this general picture
appears to hold, numerous details of how these accretion events combine to
produce a halo are unclear. The radial profile and shape of the stellar halo
are global quantities that should constrain halo formation models, since these
parameters depend on the detailed process of how individual satellite
accretions distribute their stars during the growth of the galaxy.

Extracting this information is complicated both on the observational and
theoretical side by numerous difficulties. Observationally our most detailed
measurements of the halo over wide areas of sky are often limited to volumes
within $\sim 30$ kpc of the Sun, as most are dependent on using main sequence
stars as tracers in surveys such as the Sloan Digital Sky Survey (SDSS)
\citep{juric08,dejong10,deason11}. Probing distances greater than this has
required using deep pointings over a more limited area \citep{sesar11},
intrinsically brighter but sparse tracers such as RR Lyrae
\citep{watkins09,sesar10,akhter12,cohen15}, or careful isolation of blue
horizontal branch (BHB) stars from contamination by more populous foreground
stars \citep{deason14}. These methods often suffer from either limited
sky coverage, limited number counts, or potential contaminants which are
difficult to remove. Additionally there is often the risk that the tracer
population is not representative of the entire stellar mass at that radius, and
that variations in stellar populations could produce a tracer profile that
differs from the density profile.

In this work we introduce a large survey of DDO 51 photometry that aims to
produce a well-characterized sample of giant stars in the stellar halo. The DDO
51 filter has a narrow bandpass centered on the MgH and Mg b spectral features
at $5100\rm{\AA}$, which exhibits a sensitivity to surface gravity
\citep{clark79}. \citet{geisler84} showed that the DDO 51 can be used with the
broadband Washington filter system to measure surface gravity with the M and DDO
51 filters, along with temperatures via the T$_1$ and T$_2$ filters. This was
further refined by \citet{majewski00}, who found that M - T$_2$ provided
sufficient temperature information such that the T$_1$ filter could be skipped,
thus improving the observing efficiency.

In this work we realize another gain in observing efficiency by using the SDSS
imaging for the broadband data, freeing our observing program to focus on
collecting narrow band data. Through use of the robotic United States Naval
Observatory (USNO) 1.3m telescope we have obtained 5100 square degrees of
imaging in DDO 51, yielding one of the largest samples of giants at distances
of 30-80 kpc in the Galactic halo. This enables us to measure the radial
profile and shape of the halo with greatly reduced risk that small sample
areas, varying stellar populations, or contamination are affecting our
measurements.

Our effort focuses on measuring these bulk properties of the stellar halo, the
radial profile and shape, by combining this dataset with synthetic stellar
populations models that recreate the survey footprint and DDO 51 selection. We
begin by discussing the DDO 51 data and our modeling of the resulting color-
magnitude diagram (CMD) in Section~\ref{sec_obs}. We discuss our synthetic
modeling of the survey in Section~\ref{sec_model} and our fits to the data in
Section~\ref{fits}. We finish with a discussion of other halo profile
measurements in Section~\ref{sec_discussion} and our conclusions in
Section~\ref{sec_conclusions}.

\section{Observations}
\label{sec_obs}

All observations for our ongoing imaging survey are obtained using the Array
Camera on the U.S. Naval Observatory, Flagstaff Station, 1.3 meter telescope.
The Array Camera is a 2 x 3 mosaic of 2048 x 4102 e2v CCDs with $0.6\arcsec$
pixels, and a field of view of $1.41\arcdeg$ in right ascension by
$1.05\arcdeg$ in declination. Each field is observed with a single 1200 second
exposure in the DDO51 filter. No standard star fields are observed during the
night. Fields are separated by $1\arcdeg$ in declination, and roughly
$1.34\arcdeg$ in right ascension (the separation in right ascension varies
slightly with declination, so as to cover the entire $360\arcdeg$ along a
given declination stripe with an integer number of evenly spaced fields).
Observations are taken in dark time, within one week of new moon. The
telescope is fully automated, and observations are obtained under a wide range
of seeing and cloud conditions.

\begin{figure}
\epsscale{1.0}
\resizebox{\columnwidth}{!}{\plotone{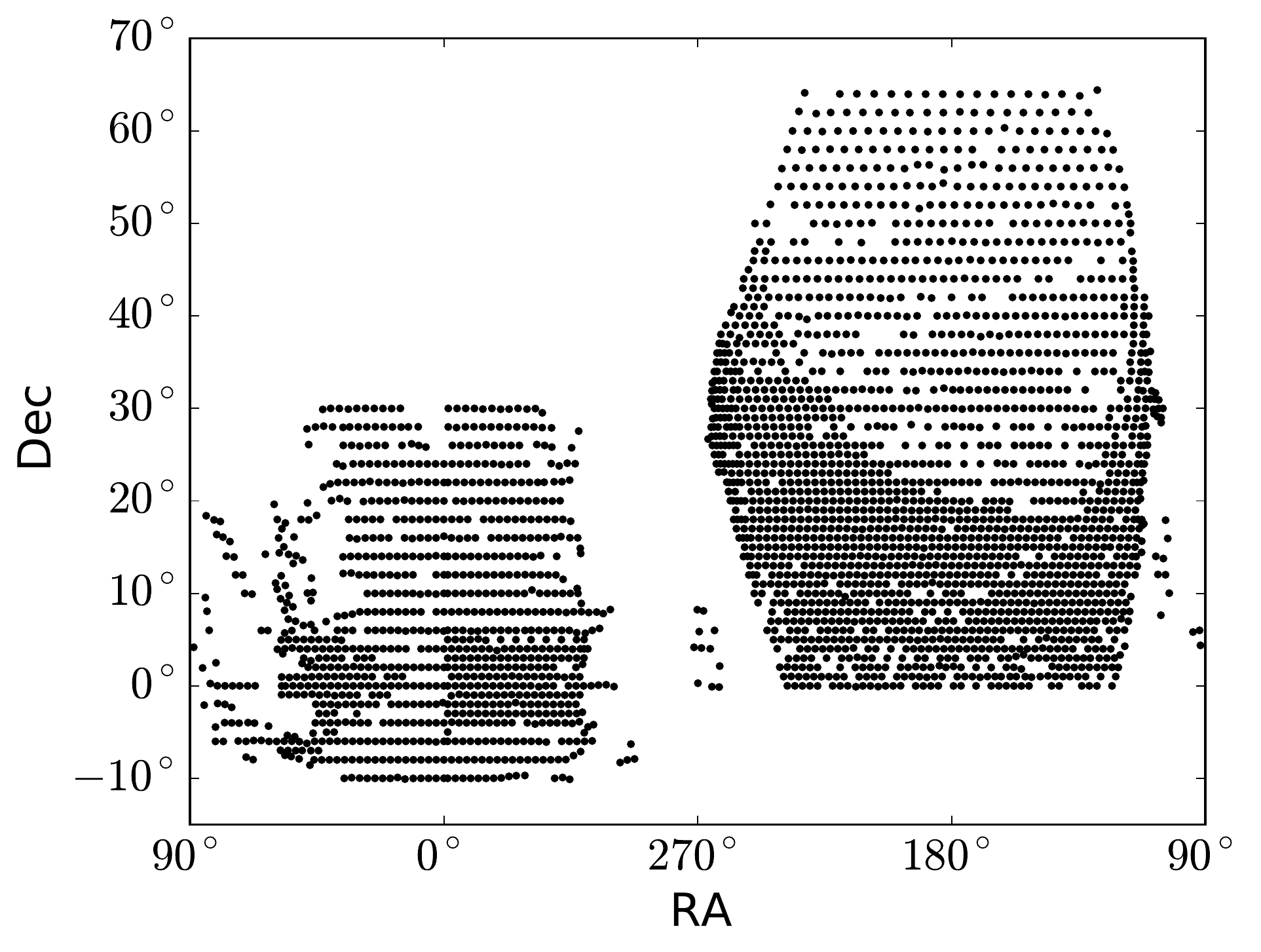}}
\caption{Sky coverage of the DDO51 observations. The survey follows the SDSS
footprint so that broadband photometry can be used from that survey. In the
northern Galactic hemisphere (right), the region at declination $<30^\circ$ is
more densely filled simply due to the ongoing survey progress.
\label{sky_coverage}}
\end{figure}

Observations commenced at the end of October 2011 and are ongoing. The analysis
in this work is based on data collected through June 2014, which includes 3,652
fields observed over 194 nights, for a total of $\sim$5113 square degrees. The
sky coverage of observations is shown in Figure~\ref{sky_coverage}. The survey
is designed to overlap the SDSS footprint so that broadband photometry is
available for all fields. This results in a survey which is focused on high
Galactic latitudes, roughly $|b| > 30^\circ$. The observation pattern first
sought to cover the SDSS footprint with a 25\% filling factor, arranged into
rows and columns in RA and Dec and skipping every other row and column. After
that was completed, we began to fill in the remaining area, resulting in a
denser coverage in some regions. The data currently cover approximately 40\% of
the target area.

Images are bias subtracted and flat-field corrected (using median twilight
flats) using the Image Reduction and Analysis Facility \citep[IRAF;][]{tody86,
tody93}\footnote{IRAF is distributed by the National Optical Astronomy
Observatories, which are operated by the Association of Universities for
Research in Astronomy, Inc., under cooperative agreement with the National
Science Foundation.}. Object detection is performed using both SExtractor
\citep{bertin96} and DAOPHOT-II \citep{stetson87}, which does a better job
detecting stars near other bright stars. DAOPHOT-II is used to model the point-
spread-function (PSF), which is allowed to vary quadratically with position in
the frame, and to measure PSF and aperture magnitudes (using a $7.8\arcsec$
radius aperture) for each object. The PSF varies considerably across the wide
field-of-view (FOV) of the telescope, introducing FOV dependent systematic
differences between the PSF and aperture magnitudes. These are fit using a
quadratic polynomial as a function of radial distance from the center of the
FOV, correcting the PSF magnitudes to uncalibrated total magnitudes. The DDO51
data are then cross-matched with SDSS to provide gri colors.

To assure the best quality data for further analysis we apply the following
cuts to the data on a chip-by-chip basis: FWHM$<$3.0\arcsec, the photometric
depth at which a signal to noise of 50 was reached must be fainter than 18th
magnitude, and RMS scatter in the polynomial fit to the $7.8\arcsec$ radius
aperture correction must be $<$0.02 mag. These criteria remove approximately
$13\%$ of all chips. All fits were also visually inspected (by D.L.N.) for
systematic or pathological issues with the data or fits. This removed 205
fields ($5\%$), mostly at low-latitude.

The unique shape of the locus of dwarf stars (``swoosh'') in the
$g-\rm{DDO51}$,$g$-$i$ diagram is used to calibrate the $g-\rm{DDO51}$ color produced from
the instrumental, aperture corrected, PSF DDO51 magnitudes. We use the data from
the entire survey to define a fiducial dwarf locus by finding the mean $g-\rm{DDO51}$
of the dwarfs as a function of $g-i$. Then to calibrate each chip, stars with
$g<19.5$ are used determine the best (i.e., lowest $\chi^2$) constant offset in
$g-\rm{DDO51}$ that brings the dwarf stars into close alignment with our fiducial
dwarf locus curve. This calibration procedure is similar to that used in
\citet{zasowski13}, who used similar data for target selection for the APOGEE
survey. This curve is inherently specific to the survey, since small differences
between DDO51 filters or between telescopes in which those filters are used
could lead to a slightly different transmission curves and thus a differing
dwarf locus. This does not impair our own analysis, but may limit direct
numerical comparison to other surveys.

\section{CMD Modeling}
\label{sec_model}

\subsection{Giant Selection}
\label{sec_giant_sel}

\begin{figure}
\epsscale{1.0}
\resizebox{\columnwidth}{!}{\plotone{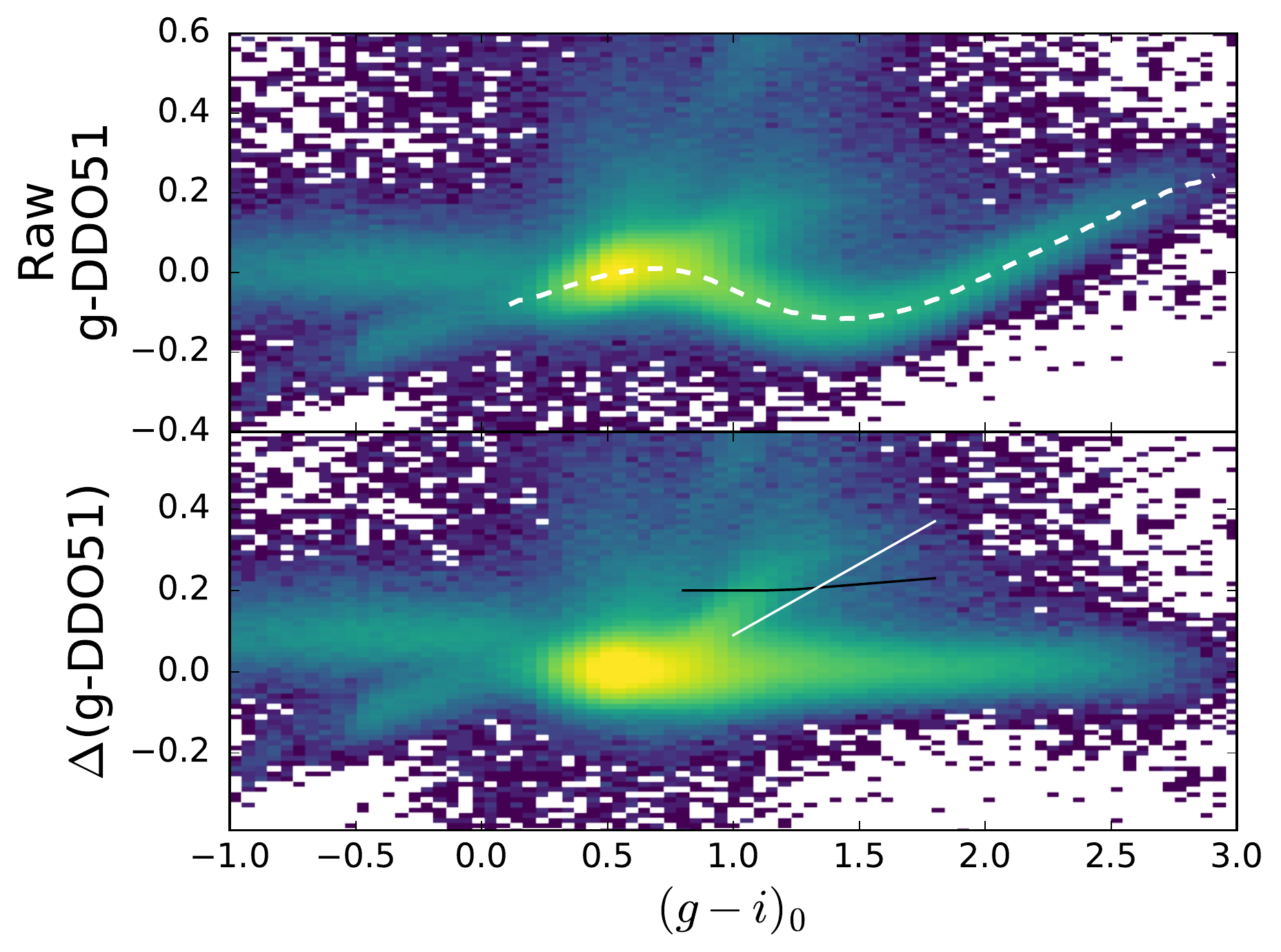}}
\caption{Color-color plots used for separating dwarfs from giants, showing all
stars in the survey. The bulk of the density is from dwarfs, but a branch can be
seen at $(g-i)_0 \sim 1$ where the giants become distinct. The top panel shows
the $g-\rm{DDO51}$ colors as observed, with the dashed line indicating the dwarf locus
``swoosh'', while in the bottom panel the g-DDO51 colors have been normalized
such that the dwarfs are centered on $g-\rm{DDO51}=0$. The giant locus can be seen
above and parallel to the white line in the lower panel. The horizontal black
line shows the color cut applied to select giants. Note that the separation
between dwarfs and giants improves at redder colors, though there are
intrinsically fewer giants towards the red.
\label{fig_sep}}
\end{figure}

\begin{figure}
\epsscale{1.0}
\resizebox{1.0\columnwidth}{!}{\plotone{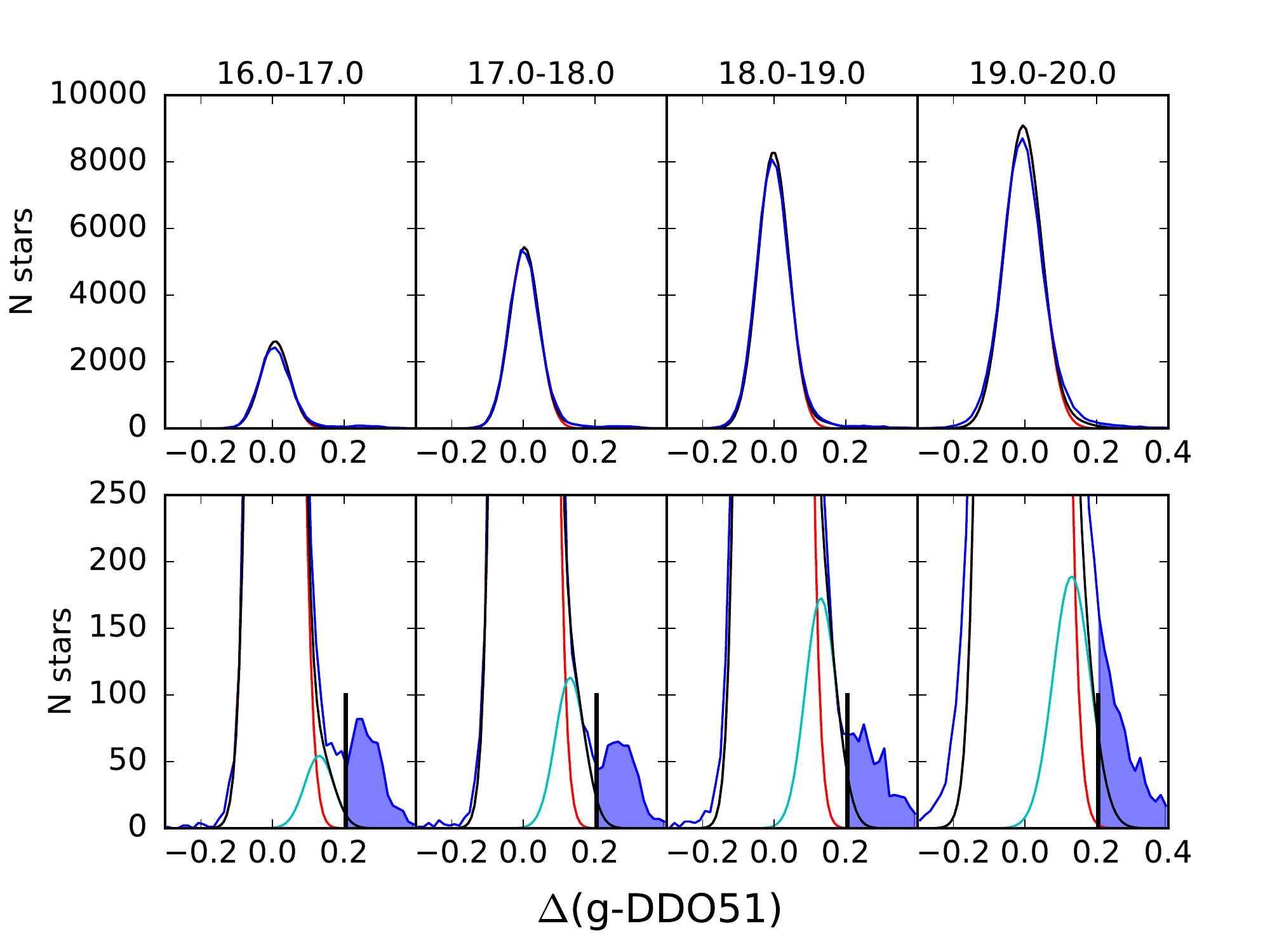}}
\caption{Histograms of \gddo for stars with $1.2 <  g-i < 1.4$. The top panels
show the full distribution in slices of apparent magnitude (bright to faint from
left to right), while the bottom panels show the same points but with the y-axis
rescaled to highlight the giant population. The observed data are in blue, our
selection for giants is marked by the vertical black line and the filled blue
area. A single Gaussian model of the dwarf contamination is shown in red, and a
model with a second offset Gaussian (cyan) added is shown by the black model
line.\label{sep_hist1}}
\end{figure}

\begin{figure}
\epsscale{1.0}
\resizebox{1.0\columnwidth}{!}{\plotone{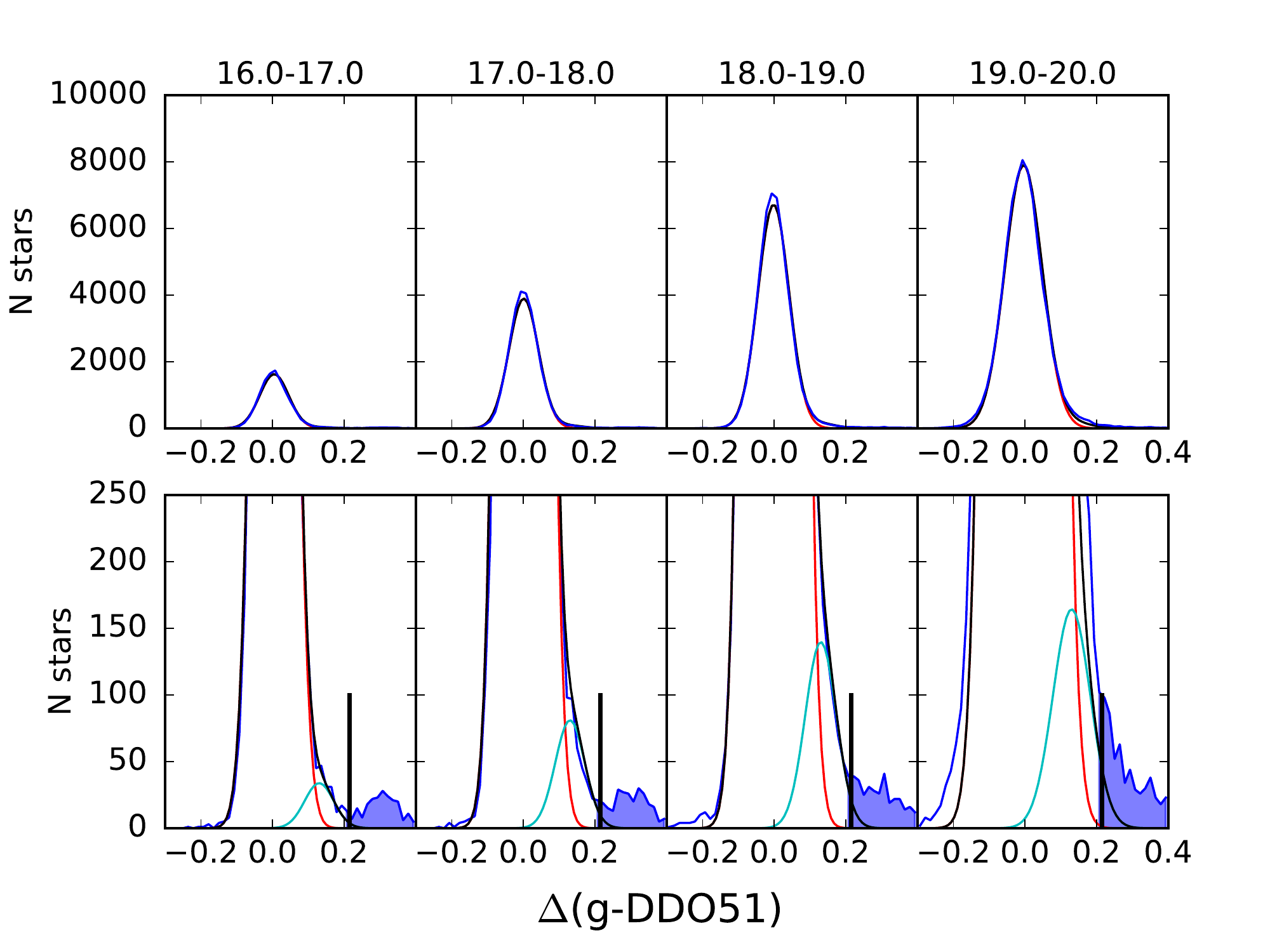}}
\caption{Same as Figure~\ref{sep_hist1}, but for stars with $1.4 < g-i < 1.6$.
The separation between giants and dwarfs is slightly better, at the cost of
diminished number counts.\label{sep_hist2}}
\end{figure}

For convenience we normalize out the ``swoosh'' feature using a polynomial fit
to the ridgeline of the dwarf locus (where giants are negligible by number),
over all fields. This produces the ``adjusted'' $g-\rm{DDO51}$ color seen in
Figure~\ref{fig_sep}, which we will refer to as \gddo.
It should be noted that the exact form of dwarf locus in
color-color space is sensitive to the particular filter's design and the
telescope in which it is used (since they are interference filters), so
specific values may be different in other DDO51 surveys.

In Figure~\ref{fig_sep} the giants lie in the feature that protrudes towards
positive \gddo{} values, starting at $g-i\sim0.8$. The best separation between
giants and dwarfs is achieved in the color range from roughly $1.2 < g - i <
1.6$, which corresponds to the upper half of the giant branch. Below that on the
giant branch (towards the blue in $g-i$) the gravity difference between giants and
dwarfs is diminished. Focusing on this target color range,
Figures~\ref{sep_hist1} and~\ref{sep_hist2} show histograms in \gddo{} of
stars with $1.2 < g-i < 1.4$ and $1.4 < g-i < 1.6$, respectively. The top row in
these figures shows the full range of the data (blue lines), and it is
immediately clear that the dwarfs dominate overwhelmingly, as expected. These
panels also include a model of what we expect from the dwarf distribution in
\gddo{} assuming that the spread is the result of photometric uncertainty plus
a small intrinsic scatter (Gaussian width $\sigma$ of 0.035 mag). For the bulk
distribution of dwarfs this works extremely well, and there is very little
deviation from a Gaussian.

Our goal, however, is to select the giants rather than modeling the dwarfs,
and these can only be seen by zooming in on the tails of this distribution.
The lower panels of Figures~\ref{sep_hist1} and~\ref{sep_hist2} are identical
to the top panels but with a different y-axis scale to highlight the giants.
This second peak in the data can be seen to the right of the dwarfs, and is
highlighted with the blue shading. We define a \gddo{} color cut (after
removing the ``swoosh'') that falls in the valley between these dwarf and giant
peaks, selecting everything to the right of that value as giants. We select as
giants all sources with corrected $\Delta(g-\rm{DDO51})>0.20$ at $g-i <
1.2$, and at $g-i > 1.2$ we use $\Delta(g-\rm{DDO51}) > (0.20 + 0.05 (g-i -
1.2))$. The slightly more aggressive selection takes advantage of the improved
separation in \gddo{} for redder giants. As will be discussed later, some dwarf
contamination is inevitable with such a method, but our modeling will seek to
account for that contamination in measuring the halo profile.

While the Gaussian fits to the dwarf distribution are overall excellent, in the
zoomed-in panels of Figures~\ref{sep_hist1} and~\ref{sep_hist2} a small
deviation between the data (blue) and the Gaussian model (red) can be seen. It
is particularly important for us to understand the source of these excess stars,
since a small fractional contamination by dwarfs could affect our inferred halo
profiles unless properly modeled. Following extensive testing with SDSS
spectroscopy, both with objects in common between datasets and with synthetic
DDO51 photometry on SDSS spectra, it has become clear that this extension to
higher \gddo{} values is from dwarfs rather than any unique class of
contaminants. It is expected that the \gddo{} distribution of dwarfs should
exhibit some variation with the metallicity of stars, since the measured
equivalent widths can clearly be diminished by either lower gravity or lower
abundance of metals. This effect is not negligible, but it is also of similar
order to the photometric scatter, such that the dwarfs with slightly higher
\gddo{} colors are not ubiquitously low metallicity but are merely on average
somewhat more likely to have lower metallicity. It is very challenging therefore
to create a complete model of the metallicity distribution function for the disk
dwarfs and their associated \gddo{} distribution. Our solution is instead to
model this heuristically. We approximate this complex metallicity function by
the addition of a second Gaussian component centered on $\Delta(g-\rm{DDO51})=0.13$ and
normalized to be 2\% of the total number of dwarfs. These parameters are fixed
across both color ranges and across magnitude bins. Figures~\ref{sep_hist1}
and~\ref{sep_hist2} show this component in cyan, and the sum of it plus the main
dwarf peak is the black line. The result matches the right-side slope of the
dwarfs much better, without any tuning specific to particular magnitude bins
(note we discard the $19 < g < 20$ magnitude bins in all fits). To be clear, we
do not interpret this as an intrinsically distinct' component centered on
$\Delta(g-\rm{DDO51})=0.13$, rather this pair of Gaussian functions is a suitable
approximation to what is likely to be a convolution of a complex MDF with
various uncertainties.

We also note that the metallicity of both the giants and dwarfs can affect their
$g-\rm{DDO51}$ color, since the abundance of Magnesium influences the
strength of the Mg b and MgH spectral features. This metallicity dependence does
not present a problem for our selection process because the lines are very weak
in giants, and thus do not show as significant of a metallicity dependence
as they do in dwarfs where the lines are strong. For dwarfs to reach the region
in color-color space occupied by giants would require metallicities of [Fe/H]=-3
or less \citep[see Figure~2 of][]{majewski00}, which is unlikely to occur in
significant numbers. These aspects combine to limit the sensitivity of our
selection process to the metallicity of either the giant or dwarf populations.

\begin{figure*}
\epsscale{1.0}
\resizebox{\textwidth}{!}{\plotone{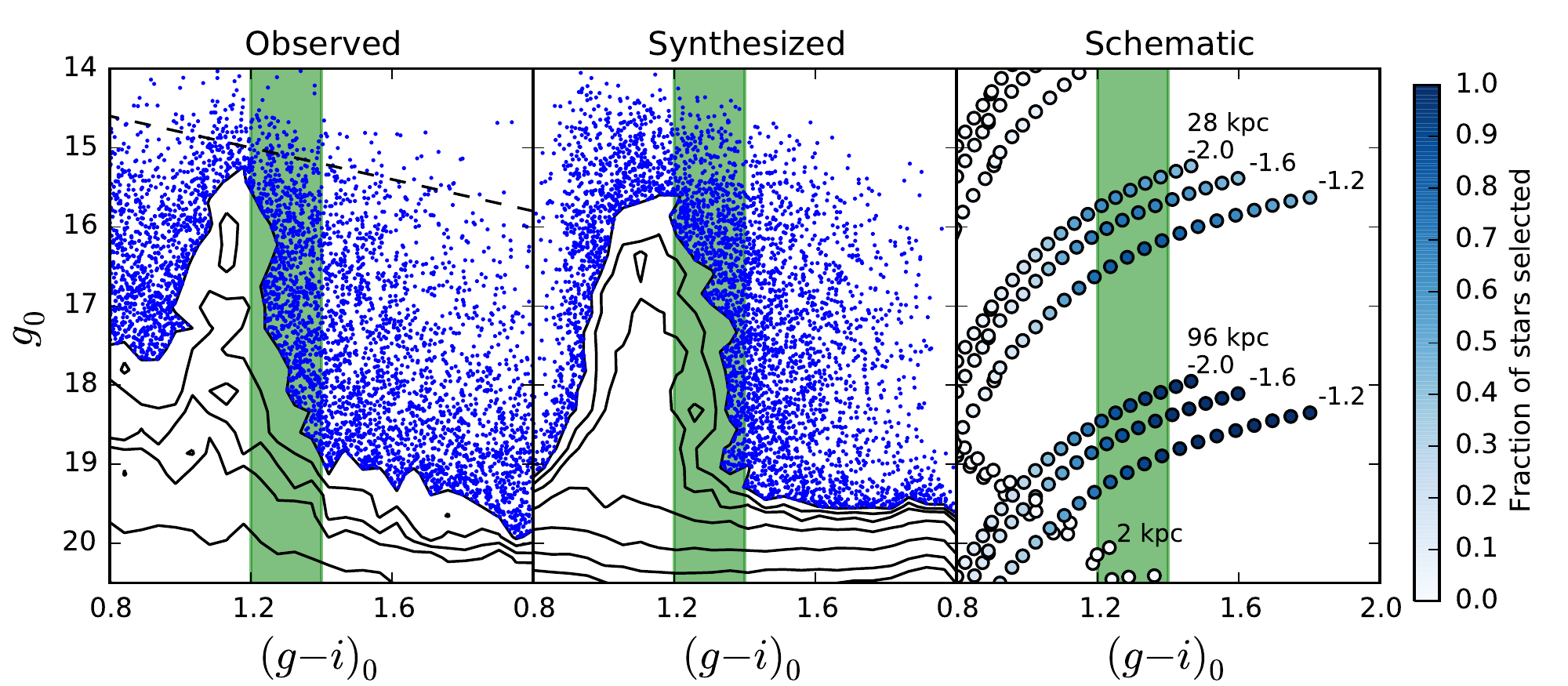}}
\caption{Observed CMD with the \gddo{} selection for giants applied (left),
along with the synthetic model CMD (center) and a schematic illustration of
the isochrones that go into the model (right). High density regions have been
replaced with contours for clarity, and the SDSS i-band saturation limit is
shown by the horizontal dashed line. The primary color range of interest ($1.2
< (g-i)_0 < 1.4$) is show in green. A discussion of the general features of
the CMD can be found in Section~\ref{qual_cmd}. Overall the synthetic model
reproduces the behavior of the observed CMD very well, including the rise of
disk contamination at $g_0 > 19.5$ and the reduced dwarf-giant separation at
$(g-i)_0 < 1.0$. The schematic diagram (right) shows the shape of the giant
branch as it appears in our data, along with the distance and metallicity
dependence (three metallicities at four distances are shown for clarity,
though in the  actual modeling we use 40 distance bins and 7 metallicity
bins). The appearance of nearby dwarfs can also be seen; fractionally they are
very well rejected from our sample, but their large numbers can cause
significant contamination at $g_0 > 19.5$.
\label{combined_cmd}}
\end{figure*}

\subsection{CMD Modeling}
%\label{sec_model}

The resulting observed CMD of selected giants is shown in the left panel of
Figure~\ref{combined_cmd}. Because the observed CMD is the result of a non-
trivial selection process, combined with Galactic density profiles of the disk
and halo, it can be difficult to interpret the resulting data by eye. We thus
sought to better understand the observed data by creating a synthetic model of
the CMD. This allows us to check our understanding of the relevant effects
that give rise to the observations, along with the sensitivity of our results
to changes in our assumed stellar populations, density distributions, and
other sources of uncertainty.

Our modeling procedure is as follows. Each line of sight of the survey is
treated independently, which are later combined in color-magnitude space to give
a CMD of the entire survey. For each line of sight, a grid of isochrones are
generated in both distance modulus and in metallicity (40 distance bins and 7
metallicity bins). For a given line of sight, any given distance modulus
corresponds to a specific Cartesian position in the Galaxy, and we can use a
Galactic density model (described below) to assign a total mass of stars we
expect to be present in that volume. We then distribute that mass among the
points on an isochrone from \citet{dotter08} according to a \citet{kroupa01}
initial mass function.

Each point of each isochrone then corresponds to a total mass at a specific
location in observed color-magnitude space. This is sufficient to determine
the fraction of stars that would pass our selection criteria at each isochrone
point, without having to generate large samples of artificial stars.
Calculating the fraction of stellar mass that meets our \gddo{} selection
criteria can be done on the basis of an isochrone point's $g-i$ color and the
photometric errors at that observed magnitude alone---a key assumption of our
model is that dwarfs and giants live on specific loci in color-color space
with only a Gaussian spread (largely due to photometric uncertainty). This
assumption works remarkably well, and allows us to compute the fraction of
stars that meet our color cut via the simple use of error functions. That is,
for each point on each isochrone, we can compute the fraction that will fall
into our $g-i$ selection window given the observed level of scatter, then
which fraction will also meet our \gddo{} selection cut.

This is then summed over all fields to yield the model luminosity
function, or without the selection cuts, can be sampled to produce a simulated
CMD. The resulting CMDs trace the increasing contamination at faint magnitudes
along with its dependence on $g-i$ color, as we will discuss below. This
process is repeated for each step in our metallicity grid independently, and
the results summed according to the modeled density in that metallicity bin.
Other stellar populations parameters are held fixed, though variation in them
could also be implemented at higher computational cost.

We parameterize the density profile of the halo as a broken power law, where both
the inner and outer power law slopes along with the break radius are potentially
variable in the fitting process (Section~\ref{fits}). This is the only density
component that is needed in the modeling process; the contamination from the
Galactic disk dwarfs is measured directly in the observations themselves (as
described at the end of Section~\ref{sec_giant_sel}). The contamination from
disk giants is assumed to be negligible over our distance range of 30 to 80 kpc.

\subsection{Qualitative CMD Properties}
\label{qual_cmd}

The resulting synthetic CMD can be seen in the center panel of
Figure~\ref{combined_cmd}, along with a schematic illustration in the right
panel. Overall, the qualitative level of agreement with the observed CMD is
excellent, and confirms that we properly understand the relevant observational
effects that play a role in generating the data. At bright magnitudes
($g<19$), the selection of giants vastly dominates over dwarfs. The central
density peak at $g-i \sim 1.1$ is caused by two competing factors: redwards of
this peak the number of observed giants is decreasing because the intrinsic
number towards the tip of the giant branch is decreasing, while bluewards of
the peak the separation between giants and dwarfs in \gddo{} becomes poorer
and fewer giants are selected observationally. The exact position of this peak
and its bluewards drop-off are thus sensitive to observational factors, and we
avoid basing our conclusions on this region. Towards the red, the
selection is much cleaner, which yields an ideal portion of the CMD from which
to measure the distant density profile of the halo. As we will show below, we
make use of two color ranges, $1.2<g-i<1.4$ and $1.4<g-i<1.6$, to further
check that our profile measurements are robust. While the $1.2-1.4$ color
range provides better statistics due to the greater number of stars, the
$1.4-1.6$ color range has a different dependence on distance and on
contamination, thus providing a double-check of our inferences.

At faint magnitudes the rise in disk contamination can be clearly seen (inside
much of the contoured region in Figure~\ref{combined_cmd} at $g=19$ and
fainter). This also exhibits a color dependence; as the dwarf-giant separation
increases towards the red, our color-cut also becomes slightly more aggressive
in rejecting dwarfs. The number of contaminating dwarfs is thus reduced, and
they do not begin to match the number of selected giants until fainter in the
CMD. Finally, at the bright end of the CMD we are limited by saturation of the
SDSS i-band data.

To match the color distribution of the observed CMD it is also necessary to
include a spread of metallicities rather than a single stellar population. As
can be seen in the schematic isochrones of Figure~\ref{combined_cmd}, if, for
example, we created a synthetic CMD with only $[\rm{Fe/H}]=-2.0$, there would be no
giants redwards of $g-i=1.5$. That would be obviously inconsistent with the
observations. Similarly, if we used only a higher metallicity isochrone, the
distribution of stars in color would be far more uniform than is observed in the
CMD, where a gradual taper in density from $g-i=1.2$ to $1.8$ is observed.
Reproducing this requires multiple isochrones of varying metallicity.
Adopting a Gaussian metallicity distribution function (MDF) of width 0.4 dex
centered between roughly $[Fe/H]= -1.5$ and $-1.8$ satisfies these qualitative
requirements and is consistent with MDF measurements in SDSS by \citet{an13}.
Figure~\ref{combined_cmd} shows that metal-poor giants are
preferentially brighter than metal-rich giants, and since our measurement range is
fixed in apparent magnitude we measure slightly further if the halo is more
metal-poor. At a metallicity of $[\rm{Fe/H}]=-1.5$ giants at
$g_0=18.5$ correspond to a heliocentric distance of 86 kpc, while
$[\rm{Fe/H}]=-1.3$ and $-1.7$ correspond to $77$ and $95$ kpc respectively.
In our fitting of the halo profile the MDF center is a free parameter, and a
further discussion of the impact of the MDF on the inferred profile can be found
in Section~\ref{metallicity}.

\section{Halo Radial Profile}
\label{fits}

\begin{figure*}
\epsscale{0.8}
\plotone{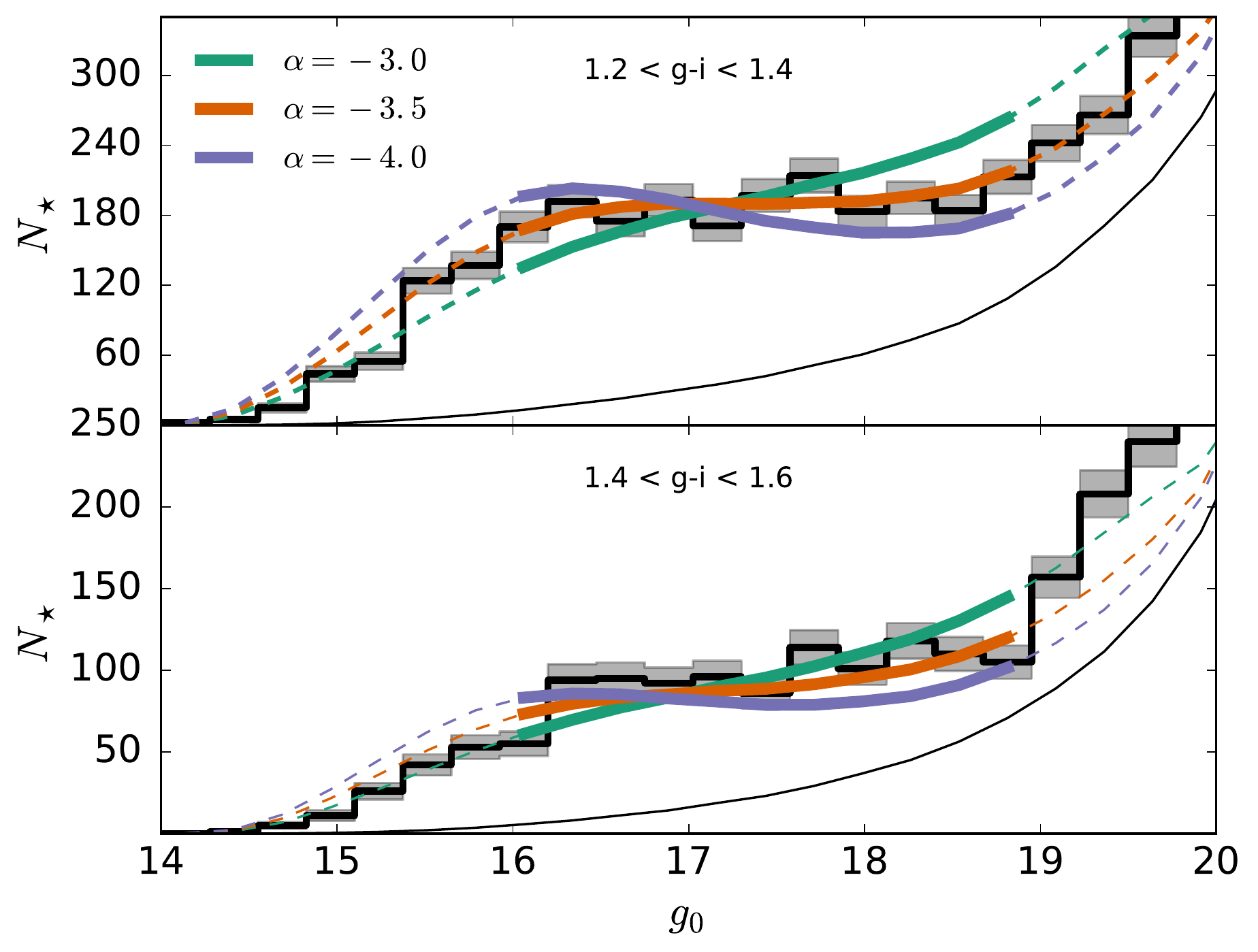}
\caption{Comparison of different outer halo slopes. In each case the halo has
a fixed $r^{-3.0}$ profile out to 27 kpc, at which point differing power law
slopes ($\alpha$) were adopted. Each model halo was then adjusted in
normalization to match the observed $1.2 < g-i <1.4$ LF, and that
normalization was carried over to the $1.4 < g-i < 1.6$ LF. Portions of the
model outside the region for fitting are shown as dashed lines, and the thin
black line shows our estimated dwarf contamination. A slope of $\alpha = -3.5$
reproduces the observations well, while steep slopes of $\alpha \leq
-4.0$ are strongly discrepant.
\label{fits1}}
\end{figure*}

Once we have confirmed that we are able to accurately create synthetic CMDs
for a given density model, we can begin to infer the best fitting spherical halo
profile. An illustration of various halo outer slopes is shown in
Figure~\ref{fits1}, with the observed data shown by the thick blue stepped
line. The estimated contamination from dwarfs is shown by the thin black
line. In each case we have normalized the model LF to match the observed data.
The upturn in counts fainter than $18.5$ is caused by contamination from disk
dwarfs. Brighter than $g_0=16$ photometric saturation in the SDSS data also
limits our observed counts, which is also modeled but should not be treated as
reliable.

Excluding these more uncertain regions, from $g_0=16$ to $g_0=18.5$ the data
show a relatively flat distribution with magnitude. While such a flat
distribution corresponds to an $r^{-3}$ density profile, in this case the
small but growing contamination at fainter magnitudes implies a slightly
steeper halo density profile of $r^{-3.5}$ after accounting for contamination.
This is readily reproduced by models with an $r^{-3.5}$ density profile, as
can be seen by eye. Models with steeper or shallower profiles at the $\pm 0.5$
level are clearly discrepant with the data, and there is no evidence of a
significant density break beyond $30$ kpc. This is true both in the $1.2 <
(g-i)_0 < 1.4$ color bin and in $1.4 < (g-i)_0 < 1.6$. With this redder color
bin we have not renormalized any of the model profiles to the redder data; the
normalization is carried over from the match to the blue color bin. The
success of our density models in reproducing both LFs simultaneously provides
an additional check that our modeling is reliable.

\subsection{MCMC Fits}

To follow up this visual comparison with a quantitative measurement, we
combine our model LF generation process with the Markov Chain Monte Carlo code
{\it emcee} \citep{foreman-mackey13}. We compare our synthetic LFs with the
data by computing the Poisson likelihood that the observed LF was drawn from
the model. Our halo density model parameters include an inner slope, a break
radius, and an outer slope, although we have limited sensitivity to the inner
portion of this profile. We include these components for the purposes of
marginalizing over them, ensuring that our outer slope measurement is not
biased or overly precise due to our incomplete data at the bright end. We
choose to only compare the luminosity functions where the data are mostly free
from bright-end saturation issues and from faint-end disk contamination, using
only the regions $16.0 < g_0 < 18.5$. Both the inner slope and the outer slope
were constrained in the MCMC sampling to values between $-2.0$ and $-4.0$, since
by visual inspection the outer slope was clearly well within this range and
literature measurements of the inner slope had found similar values \citep[See
Table 5 in][for reference]{akhter12}. The break radius was constrained to
within 15 and 45 kpc, again bracketing the range of measurements from previous
works. The addition of a Gaussian prior on the break radius had little impact
on the resulting outer slope measurement except to reduce the uncertainties.
The halo MDF was kept constant in the fit, since measuring the MDF is beyond
the scope of this work.

The resulting measured outer slope value is $\alpha = -3.5 \pm 0.2$, which is
clearly in agreement with the visual comparison in Figure~\ref{fits1}. There is
very little covariance between the outer slope and the inner slope or break
radius; our data simply do not probe these regions sufficiently.

\begin{figure}
\epsscale{1.2}
\resizebox{\columnwidth}{!}{\plotone{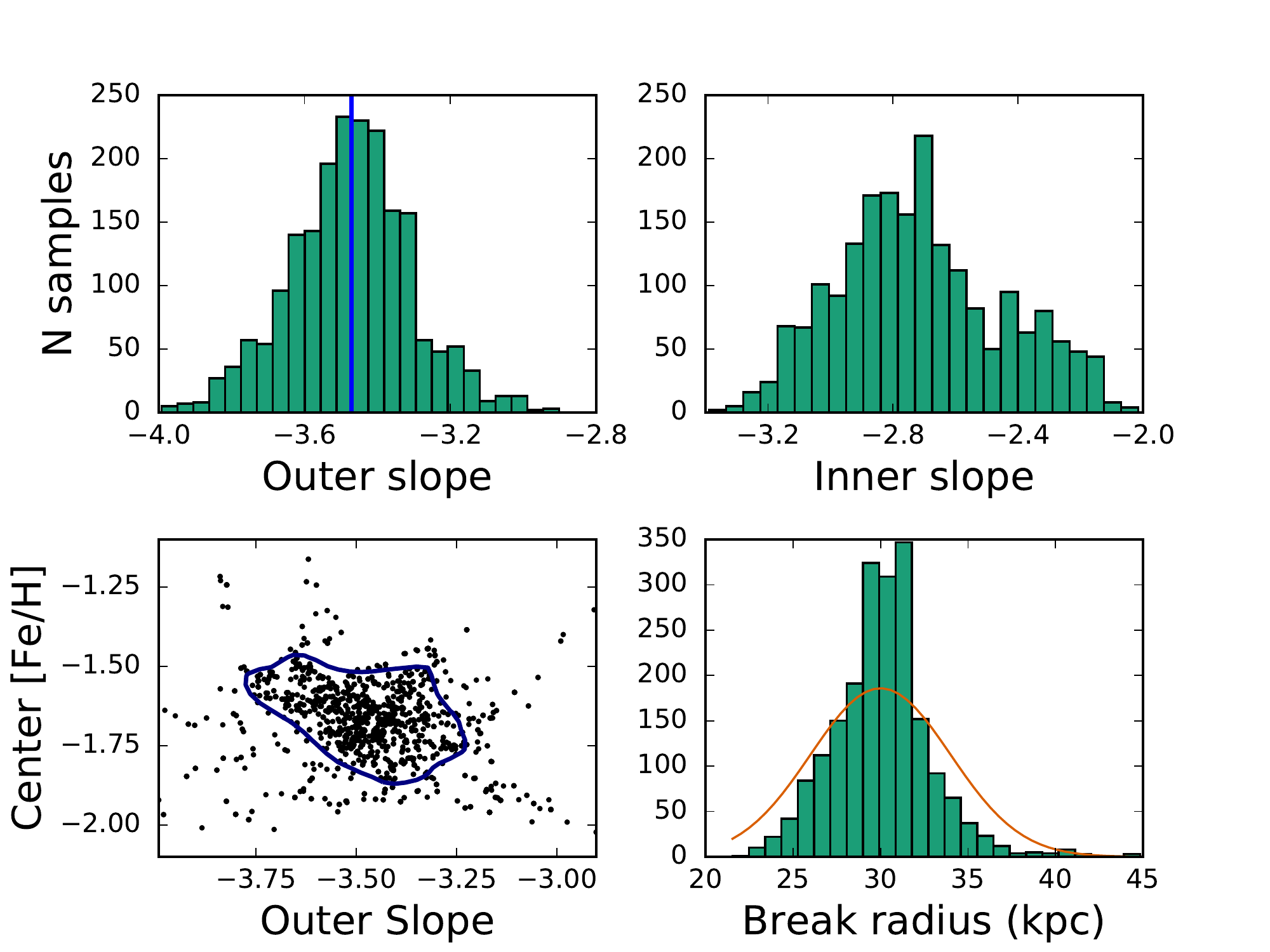}}
\caption{Marginalized probability distribution functions for three shape
parameters, and MCMC samples showing the covariance between metallicity and
outer halo slope (bottom left, $1\sigma$ area outlined in blue). The break
radius prior is shown in orange, while the two slopes had flat priors. The break
radius is not strongly constrained by this fit.
\label{mcmc_params}}
\end{figure}

The marginalized probability distribution functions for the halo shape
parameters can be seen in Figure~\ref{mcmc_params}. While the outer slope is
well-constrained, the inner slope fit permits a wide range of values since we
have little data in this region. The break radius is only slightly more
constrained than the imposed prior (shown in orange). We also show the
covariance between the outer slope measurement and the halo metallicity center.

\begin{figure}
\epsscale{1.2}
\resizebox{\columnwidth}{!}{\plotone{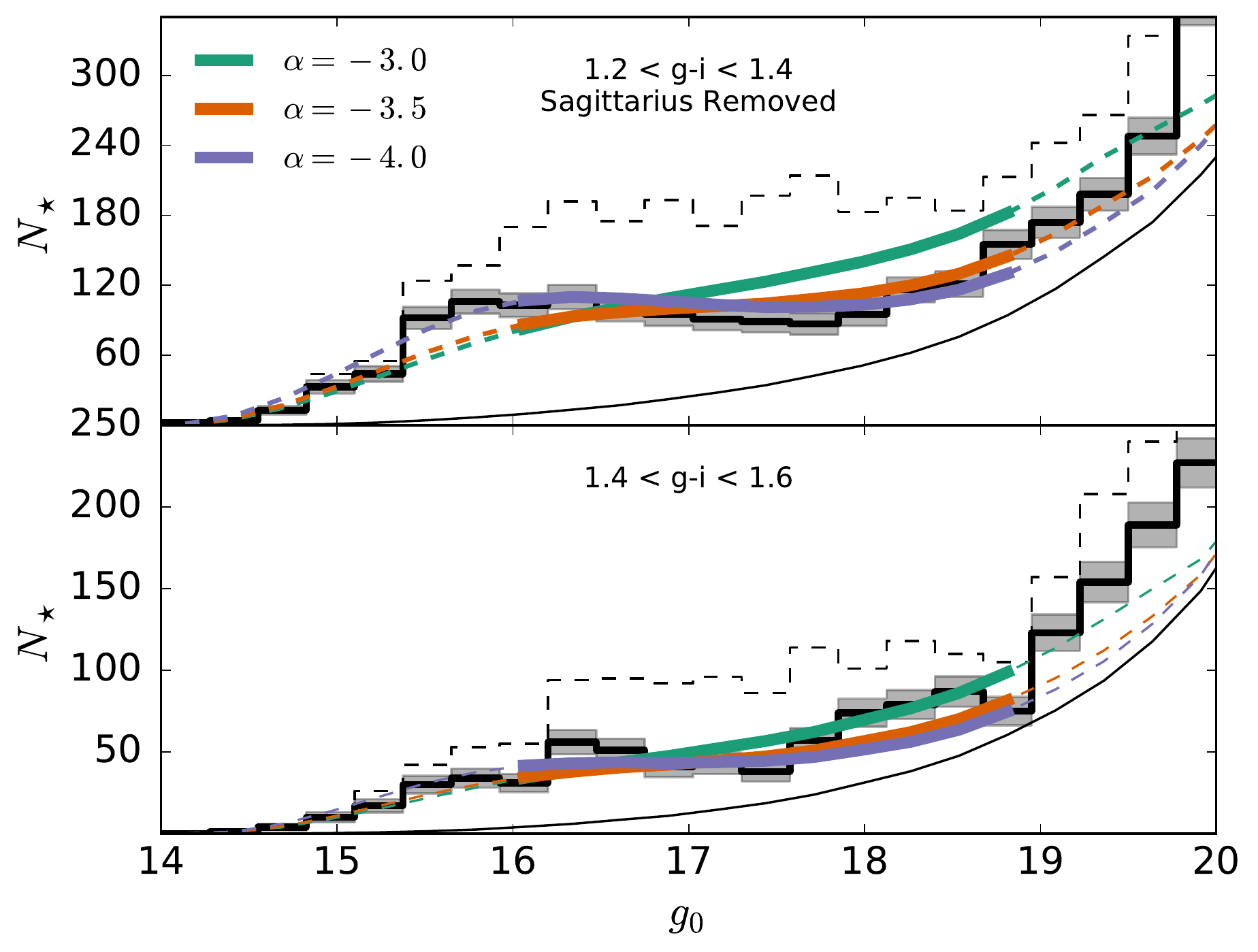}}
\caption{Observed luminosity functions with all fields within $12^\circ$ of
the Sagittarius plane removed. The original LF is shown as the dashed blue
histogram, and range of model slopes are overplotted (similar to
Figure~\ref{fits1}). The Sagittarius stream is not driving our halo radial
profile measurement. \label{sgr_removed}}
\end{figure}

The interpretation of a single radial profile as having broad significance for
the entire halo is complicated by the existence of, and likely dominance of,
accreted and poorly-mixed substructure in the halo. This presents the question
of whether one is able to measure a halo-wide quantity based on a limited number
of pointings with varying degrees of identifiable substructures present in
each? Clearly we would not put much confidence that a single line of sight
through the halo, which would result in a radial profile which rises and falls
as it passes through various structures, would be strongly representative of
the entire halo. Our wide survey footprint is well-suited to avoiding such an
effect, but there remains a risk that, because the Sagittarius stream covers a
large fraction of this footprint, our radial profile may be more a measurement
of Sagittarius than of the halo.

To mitigate this concern, we show in Figure~\ref{sgr_removed} a luminosity
function where all fields within $\pm12^\circ$ of the Sagittarius plane
\citep[as defined by][]{majewski03} are excised from the data. The resulting
contribution of the halo relative to the disk is reduced due to the decreased
coverage at high Galactic latitudes, but the inferred radial profile is
essentially unchanged. In part this is due to the mass of Sagittarius not
dominating the total mass of the halo over the region in which we are
sensitive, and is also a result of the significant distance variation that
Sagittarius exhibits; it spreads material across the LF rather than
piling up at at a single distance. Performing our MCMC fits on the
Sagittarius-excised data produces a halo slope of $\alpha=-3.5 \pm 0.2$,
indistinguishable from the fit with the entire dataset. Because
Sagittarius is the most dominant known halo substructure, we argue that our
measured radial profile is not dominated by any single structure, and can be
meaningfully (but imperfectly) representative of the halo in bulk.

\subsection{Metallicity Effects}
\label{metallicity}

\begin{figure}
\epsscale{1.2}
\resizebox{\columnwidth}{!}{\plotone{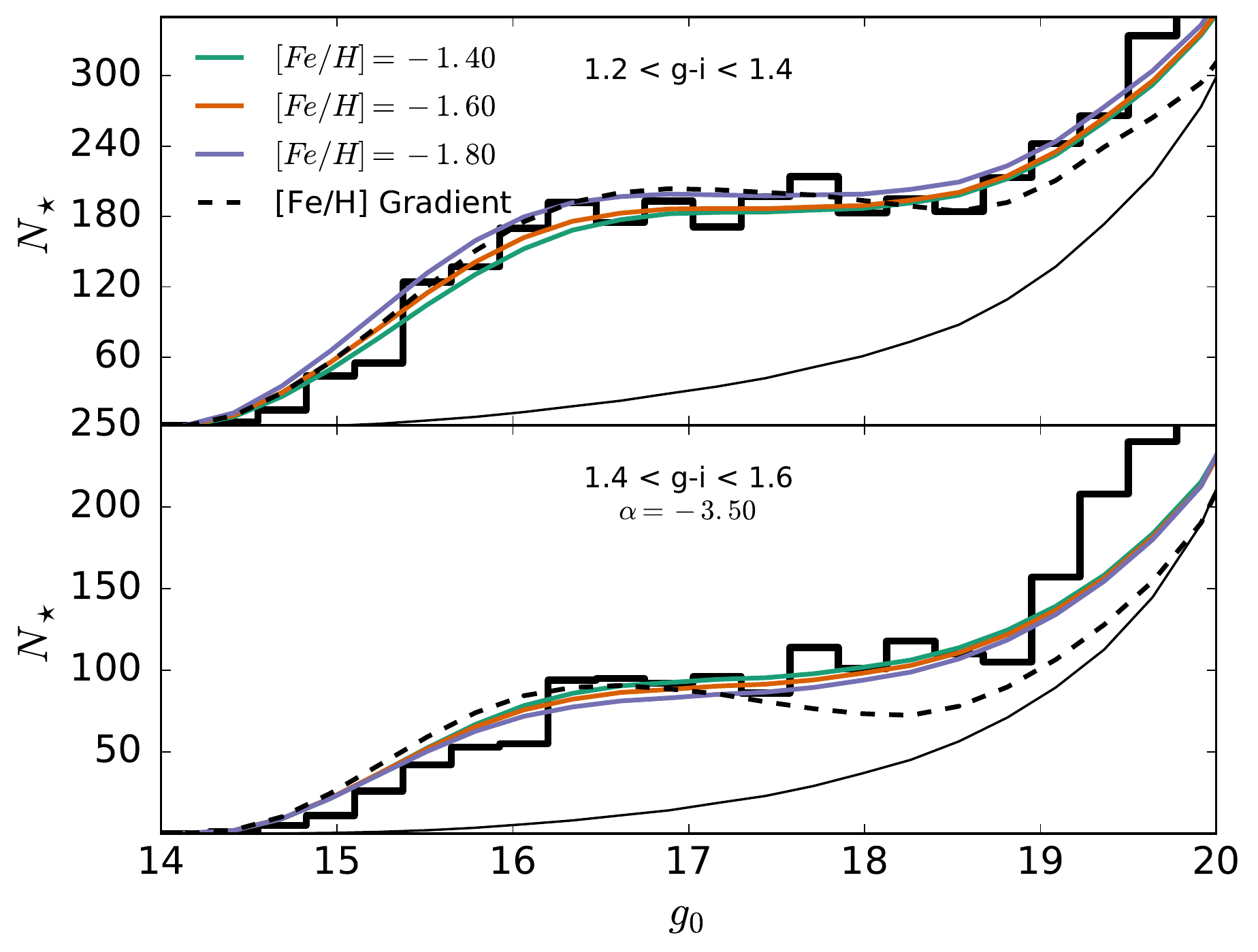}}
\caption{Metallicity dependence of the synthetic LF, for both color
selections. Each model has an $r^{-3.5}$ outer density slope but different
mean metallicities.
Different metallicities impact the slope slightly due to the different
distance ranges they cover in the halo. More notably, different assumed
metallicities impact the relative counts between the two color bins, as a more
metal-poor halo tends to push stars into the bluer color bin.
An artificially steep metallicity gradient is
also shown by the dashed black line, which strongly disagrees with the data.
\label{fits3}}
\end{figure}

Our analysis also depends on the metallicity distribution function (MDF)
for the halo. As discussed above, the metallicity of the halo determines what
portion of the giant branch falls into our selected color cuts, and also sets
the distances we infer for these stars. We discussed above that a broad MDF is
required both by studies of the local halo stars and by the color distribution
of giants in our sample. However, the center of this MDF is uncertain,
particularly at the large distances probed by our sample of giants.
Varying the MDF has two primary effects on our models: a different
metallicity profile changes where the $\sim 30$ kpc density break appears in
magnitude space, and it also changes the relative number of counts between our
two color bins. These effects can both be seen in Figure~\ref{fits3}. To
accommodate these effects we have left the center of the MDF as a free parameter
in the MCMC fitting, with a fixed width of $0.4$ dex. The resulting metallicity
preferred by the fit is [Fe/H] = $-1.7 \pm 0.14$. This falls in between the halo
metallicity reported by \citet{an13} of their ``coadd sample'', [Fe/H] =
$-1.5$, and their ``calibration sample'', [Fe/H] = $-1.8$. The value from our
fit is reasonable given the state of independent measures of the halo
metallicity. The impact of this metallicity fit on our inferred halo profile is
shown in the lower left panel of Figure~\ref{mcmc_params}. While changing halo
metallicity can drive the outer slope measurement by a few tenths, any
metallicity that can reasonably fit the fraction of stars in the two color bins
still produces a halo profile slope between $-3.75$ and $-3.25$.

A scenario which does potentially present a greater impact on our inferred
profiles is that of a halo metallicity gradient. In this case a different
portion of the giant branch at each distance will fall into our color cut, and
can alter the observed number counts. While some previous studies, such as
\citet{dejong10}, have found decreasing metallicities between 10 and 30 kpc
from the sun, other studies have argued for a constant metallicity in this
region \citep[e.g.][]{ivezic08}. In either case, we can evaluate the effect
that such a scenario would have on our inferred density profiles. To do this
we created a toy metallicity gradient model, where the MDF center changes
linearly from [Fe/H]$= -1.3$ at 20 kpc (Galactocentric) to [Fe/H]$ = -2.0$ at
100 kpc. The MDF width of $0.4$ dex is preserved at all radii. The result of
this model can be seen in Figure~\ref{fits3} by the dashed black line, where
the underlying density profile is $\alpha=-3.5$. The result is an artificially
steep drop in the LF, despite having the same density profile. A negative
metallicity gradient would cause us to infer an artificially {\it steep}
density profile. This can be understood intuitively by comparison to the
isochrones in Figure~\ref{combined_cmd}. At bright magnitudes the high
metallicity isochrone dominates, and the $1.2 < g-i < 1.4$ color cut samples
relatively faint on the giant branch. Moving to larger distances and lower
metallicities puts the top of the giant branch into the color cut, which has
significantly fewer stars due to the intrinsic steepness of the giant branch
luminosity function.

It is clear then that only a positive metallicity gradient would cause us to
erroneously infer a shallower density profile than was actually present, but
there is presently no evidence for such a gradient in the halo.

\subsection{Halo Oblateness}

\begin{figure*}
\epsscale{1.0}
\plotone{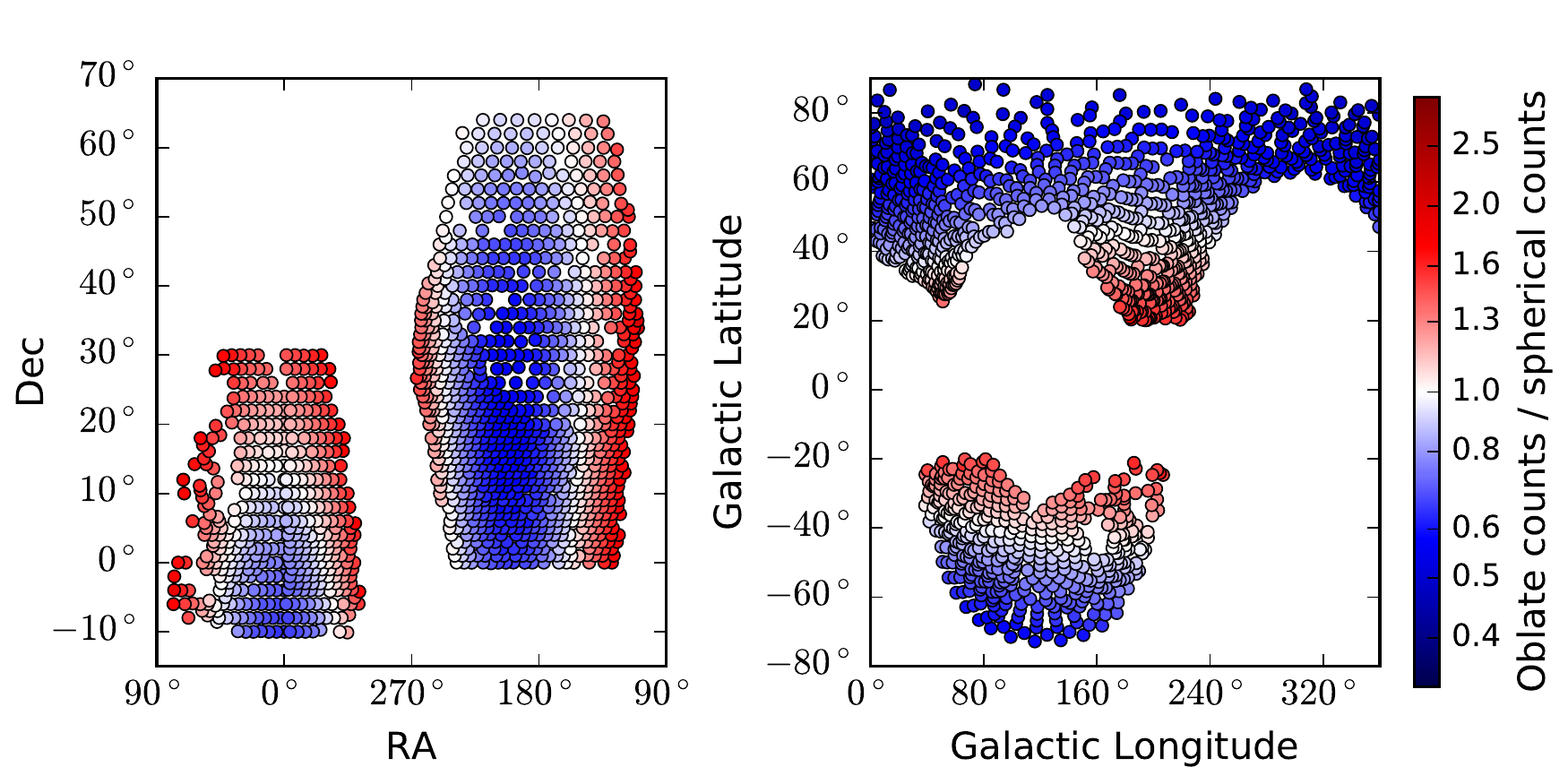}
\caption{Ratio of counts in a $q=0.6$ flattened model to a spherical model,
shown in equatorial coordinates on the left and Galactic coordinates on the
right. This illustrates the form of the effect we would see in a non-spherical
halo, with mostly a dependence on Galactic latitude but also slightly
modulated by Galactic longitude (since the Sun is not in the center of the
Galaxy).
\label{oblateness_model}}
\end{figure*}

\begin{figure}
\epsscale{1.0}
\resizebox{\columnwidth}{!}{\plotone{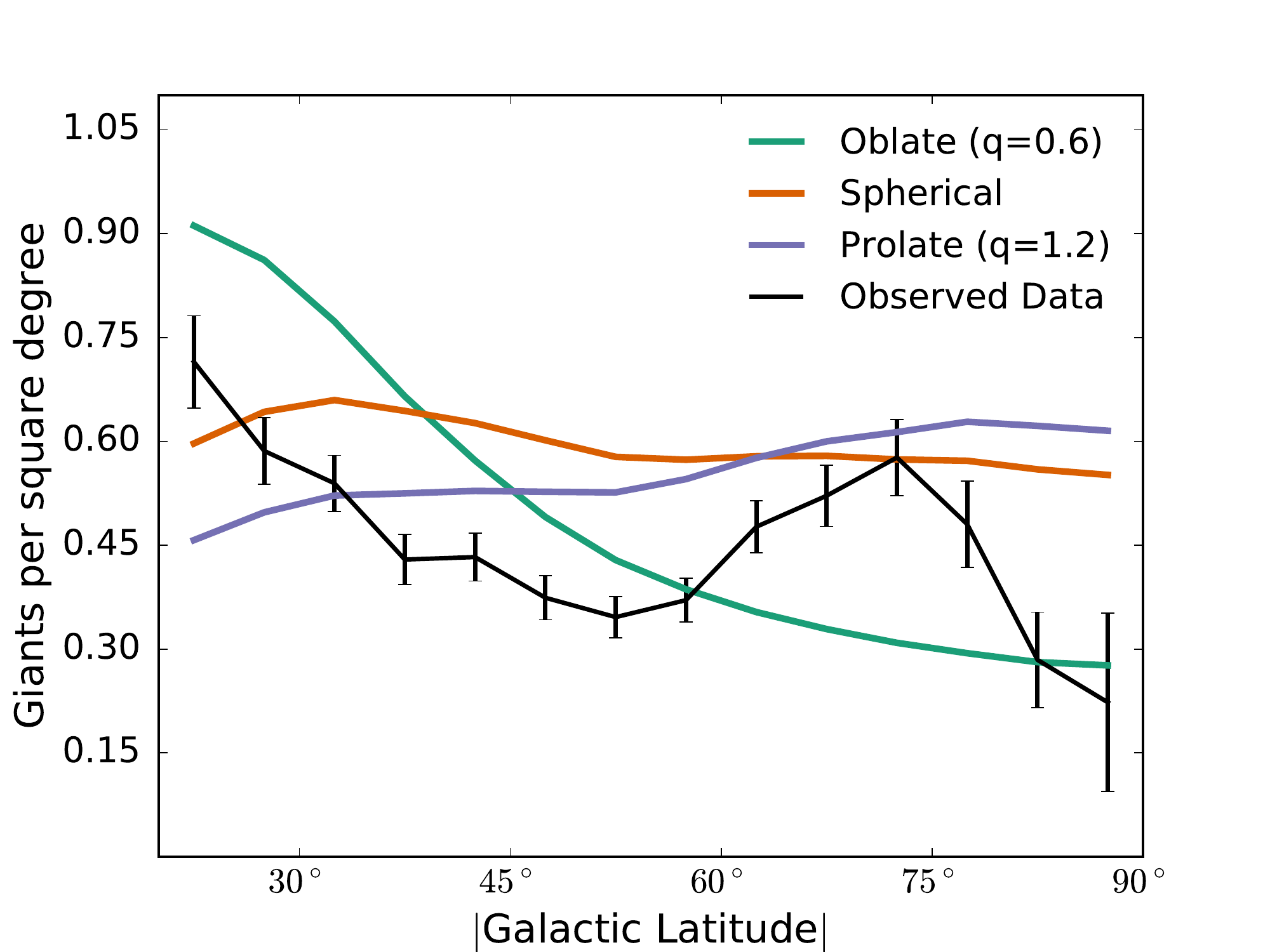}}
\caption{Visualization of the halo shape measurement. The mean number of giants
per square degree of surveyed area is computed (black), and compared to the
expectations for flattened ($q=0.6$, blue) and prolate ($q=1.2$, red) models and
a spherical halo model (green line). The data exhibit an excess at both high and
low latitudes such that no ellipsoidal shape could be considered a good fit.
\label{latitude_plot}}
\end{figure}

\begin{figure}
\epsscale{1.0}
\resizebox{\columnwidth}{!}{\plotone{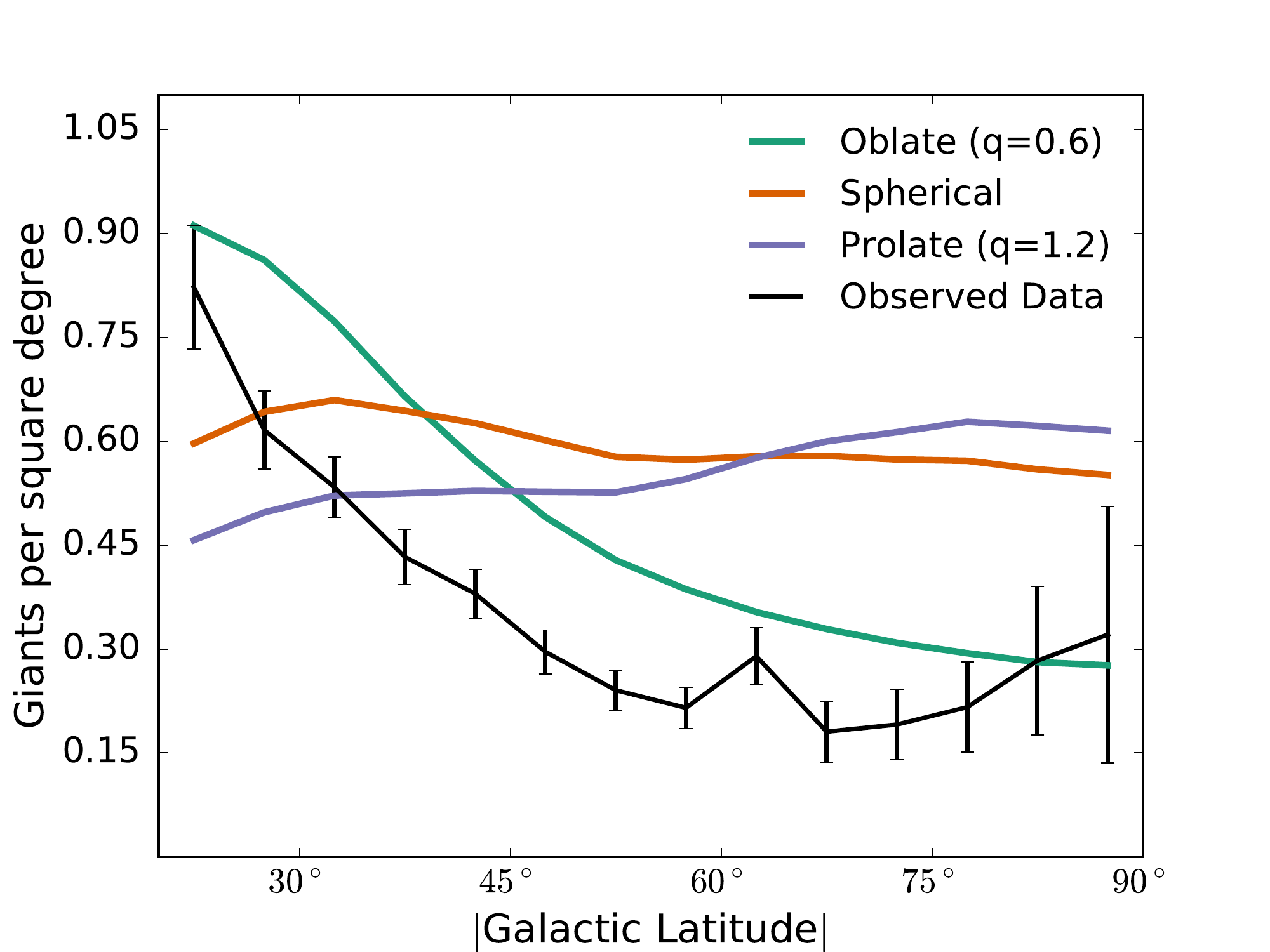}}
\caption{Halo shape models after removing all fields within $12^\circ$ of the
Sagittarius stream. The models are not renormalized to fit to the subselected
data. This behavior of this subset of the data is more consistent with an oblate
halo, suggesting by comparison with Figure~\ref{latitude_plot} that the
Sagittarius stream was responsible for an overdensity of stars at high latitude.
\label{latitude_plot_nosgr}}
\end{figure}

In addition to the radial profile, the shape of the stellar halo as a whole
may also hold information about the formation mechanisms of the halo and
potentially about the accretion history of the Milky Way. This is particularly
true in light of the suggestion that some component of the halo may be the
result of ``kicked up'' Galactic disk stars, which are scattered to large
heights by mergers \citep{zolotov09,purcell10,font11}. Such stars may leave a
signature in the resulting shape of halo, as one might expect a flattened
distribution of stars if their orbits have not been completely randomized. In
models such as these the kicked-up component is only significant in the inner
10-30 kpc of the halo, and distances beyond this are dominated by accreted
stars. Our data sample this outer region where the kicked up disk star
fraction is likely to be negligible, thus a shape measurement at these
distances presents an informative comparison to similar measurements of the
inner regions of the halo.

Our procedure for measuring the halo shape is to construct model LFs for
each telescope pointing individually, rather than a single LF for the sum total
of the survey. These individual LFs will vary in their overall level depending
on their line of sight through their halo. An illustration of the magnitude we
expect to see is shown in Figure~\ref{oblateness_model}, which shows the ratio
of counts expected in a model with $q=0.6$ over a spherical model. As expected,
this ratio is mostly dependent on Galactic latitude, and much of the information
in the measurement comes from low latitude fields towards the Galactic
anticenter and in the southern Galactic hemisphere. It is important to note that
the shape of the luminosity function does not change with varying oblateness; an
$r^{-3}$ halo still produces a constant LF with magnitude regardless of
flattening. For this reason we use only our best fit radial profile from
Section~\ref{fits} but with varying flattening factors. Changing the flattening
of a model halo at fixed normalization also changes the total number of stars it
predicts in our survey, hence in discussing non-spherical models we also adjust
the normalization so that it matches in number the overall LF.

A visualization of the halo shape is shown in Figure~\ref{latitude_plot},
showing the surface density of giants against Galactic latitude. Model halos are
also shown for comparison, both flattened, spherical, and prolate.
It is clear that no ellipsoidal halo model is a suitable
fit. There is enhancement in density at both low and high latitudes with a
depression in between, which demands a degree of freedom not allowed by either
oblate or prolate models. Attempting to fit this with such a model would not
produce any acceptable summary of the halo's shape.

As can be seen qualitatively in images like the ``Field of Streams''
\citep{belokurov06} and quantitatively in measurements such as \citet{bell08},
\citet{starkenburg09}, and \citet{xue11}, the halo contains significant density
substructure which may make spherical or ellipsoidal models poor fits to the
data. To illustrate the effect of substructure in our data,
Figure~\ref{latitude_plot_nosgr} shows the same plot of the density with
Galactic latitude, but where all fields within $12^\circ$ of the Sagittarius
plane have been excluded. The remaining fields are broadly consistent with a
flattened halo, and shows a significantly reduced density at high latitude
compared to the full sample.

One could be tempted to consider a measurement with Sagittarius excluded as
more representative of the halo, but this implies an unphysical distinction
between massive structures of tidal debris that are recognizable and the rest
of the substructured halo which does not appear as individually recognizable
features. We do not report a fit of the Sagittarius-removed data for this
reason. A summary statistic of the halo shape that accounts for and captures
the prevalence of substructure would be a much preferable treatment, but this
is a complex task which we leave to a subsequent work.

\section{Discussion}
\label{sec_discussion}

The stellar halo beyond $30-40$ kpc is a difficult region to study; tracers
bright enough to be found robustly at the depth of current large-scale surveys
are rare and can be difficult to distinguish from nearby contaminants. The
result of this is that studies have typically used either RR Lyrae or BHB
stars for this region where main-sequence stars are not readily accessible.

Studies of RR Lyrae have reported steep slopes of $\sim-4$ to $-5$ beyond
$20-30$ kpc, e.g., $-4.5$ by \citet{watkins09}. The region of sensitivity is
comparable to our work, yet significantly discrepant from our measurement of
$-3.5$. The origin of this disagreement is unclear, but we can consider a few
possible sources. The \citet{watkins09} work uses RR Lyrae from the SDSS stripe
82, which covers 290 deg$^2$. It is possible that slopes measured in
relatively small patches of the sky will be highly variable due to the
substructured nature of the halo, as surveys may happen upon underdense or
overdense regions. This is independent of the tracer used, and can only be
alleviated with larger surveys.

A survey of BHB stars in the 2dF Quasar Redshift Survey by \citet{depropris10}
over a $750$ deg$^2$ area and with data extending out to 90 kpc found a much
shallower slope of $-2.5$. A much smaller survey used deep observations of F
stars over a $23$ deg$^2$ field to obtain a slope of $-4.85$ to 60 kpc
\citep{piladiez15}. The same caveat of substructure applies in these surveys
as well; we expect large variations in halo densities measured by smaller
surveys. We note that our sample may be similarly affected, if at a lesser
degree due to the larger area coverage. The substructured halo will cause
variation between surveys above and beyond Poissonian uncertainties.

One of the larger samples of halo stars beyond 30 kpc comes from
\citet{deason14}, who used a photometric selection process to identify BHB
stars in the SDSS imaging footprint. While this does alleviate much of the
issues with smaller surveys, the photometric selection of BHB stars is a
particularly fraught process. We note that at magnitudes needed to probe the
outer halo with BHB stars the number of contaminant blue stragglers and
quasars substantially exceeds the number of target BHB stars \citep[see Fig. 8
of][]{deason14}; this is the case for most tracers of the outer halo. The
separation in broadband color between the target BHB population and the
contaminant population is, however, only $0.09$ magnitudes according to the loci
assumed by \citet{deason14}, though the intrinsic widths of the
distributions, prior to any photometric error, are $0.04$ magnitudes. This
separation becomes even more difficult in the presence of the photometric
uncertainties at $g \sim 19$. The lack of decisive separation between the
populations presents a tremendous challenge to any attempts at extracting the
individual densities, and makes the resulting density profile particularly
sensitive to any discrepancies between the chosen models and the behavior of
the actual populations.

We argue that our data have a much clearer separation between the target
tracer population and any contaminants, though at the cost of much additional
telescope time in obtaining the narrowband data. Our work significantly
disagrees with the finding of a halo slope of $-6$ beyond 50 kpc by
\citet{deason14}. Such a steep profile would require essentially zero giants
to exist between 18th and 19th magnitude in our survey, and less than half of
the number that we detect between 17th and 18th magnitudes (see the middle
panels of Figure~\ref{sep_hist1}).

Another program that has sought to characterize the halo profile is presented
in \citet{xue15}, where SEGUE spectroscopy provides the additional information
for separating bright giant tracers from contaminants. Of their fitted models, their
broken power law model is most similar to ours since our data start beyond
their break radius, and they find a power law slope of $-3.8 \pm 0.1$. This is
similar to our result, though slightly steeper. We note though that this
measurement is primarily driven by data inside of $\sim 50$ kpc; beyond this
distance giants become rare in the SEGUE selection (a total of seven stars
beyond 65 kpc) and Poisson noise becomes significant \citep[See
Figure~7,][]{xue15}. Even with this small number of stars, \citet{xue15} find
their data to be formally inconsistent with the steep drop beyond $50$ kpc
suggested by \citet{deason14}. Further confirmation of this comes from the RR
Lyrae profile of \citet{cohen15}, who find a profile consistent with $r^{-3.8}$
between 50 and 100 kpc. Their sample is similar to the spectroscopic giant
measurements of \citet{xue15} in that the sample is extremely clean, but with
the risk of incompleteness at large distances which must be corrected
statistically.

\section{Conclusions}
\label{sec_conclusions}

Our results show an extensive and unbroken halo beyond $\sim 30$ kpc out to
$80-90$ kpc, with a power law profile of $-3.5$. Our ability to measure the
profile robustly at these large distances is a result of bringing additional
decisive information beyond broadband photometry, enabling a strong separation
between the intrinsically bright tracers of interest and the more numerous dwarf
star contamination. Our data cover a large area of sky to significant depth,
further mitigating the common problems of sparse sampling and biases towards
nearby regions.

While knowledge of our own Galaxy is of interest on its own, our derived
Milky Way halo density profile will be particularly informative in
comparison with both other galaxies and with simulations and models of stellar
halo formation in general. We are only recently able to start putting our own
outer halo in context with those of a number of nearby galaxies, e.g., NGC 253
\citep{bailin11,greggio14}, M 101 \citep{vandokkum14}, Andromeda
\citep{courteau11,ibata14}, and the several galaxies in the GHOSTS survey
\citep{monachesi15}. These comparisons will give us an observational
understanding of the degree to which the Milky Way's halo is ``average'' or an
outlier, and will begin to tells us about the accretion history of the Milky
Way in general. This is part of what motivates our focus on the entire halo
rather than removing ``substructure''; this substructure (or lack thereof) is
likely what carries the information on the accretion history that we are most
interested in. Comparing the bulk properties of the halo, rather than the
properties after removing substructure, is the only level basis for comparison
we will have between the Milky Way and external galaxies. Any identification
and removal of individual structures is closely tied to the specific types of
observational data available, and the character of available data is vastly
different between external galaxies and our own.

Pursuing these comparisons in depth will require further work, both
observational and theoretical. Theoretically, we must develop a robust
understanding of the connection between a stellar halo profile and its
underlying dark matter halo and accretion history. This will require high
fidelity simulations of diverse halo assembly histories, though this is a
challenging area for simulations \citep{bailin14}. Observationally, a full
comparison between halos will require metrics that can account for and measure
substructure quantitatively; this will be the subject of a subsequent work.

\acknowledgments

This work was partially supported by NSF grant AST 1008342. D.L.N. was supported
by a McLaughlin Fellowship at the University of Michigan. S.R.M. was supported
by NSF grant 1413269.

Funding for SDSS-III has been provided by the Alfred P. Sloan
Foundation, the Participating Institutions, the National Science
Foundation, and the U.S. Department of Energy Office of Science. The
SDSS-III web site is http://www.sdss3.org/.
SDSS-III is managed by the Astrophysical Research Consortium for the
Participating Institutions of the SDSS-III Collaboration including the
University of Arizona, the Brazilian Participation Group, Brookhaven
National Laboratory, Carnegie Mellon University, University of
Florida, the French Participation Group, the German Participation
Group, Harvard University, the Instituto de Astrofisica de Canarias,
the Michigan State/Notre Dame/JINA Participation Group, Johns Hopkins
University, Lawrence Berkeley National Laboratory, Max Planck
Institute for Astrophysics, Max Planck Institute for Extraterrestrial
Physics, New Mexico State University, New York University, Ohio State
University, Pennsylvania State University, University of Portsmouth,
Princeton University, the Spanish Participation Group, University of
Tokyo, University of Utah, Vanderbilt University, University of
Virginia, University of Washington, and Yale University.

\end{document}